\documentclass[a4paper,12pt]{article}
\begin{document}
\baselineskip=21pt
\title{Non-canonical Quantization of a Quadratic Constrained
 System\footnote{submitted to Phys. Rev. Lett.}}
\author{Metin Arik and G{\"o}khan {\"U}nel}
\date{Bo{\~g}azi{\c c}i University, Centre for Turkish-Balkan Physics Research and Applications,
        {\. I}stanbul, Turkey}
\maketitle

\begin{abstract}
We propose an alternative to Dirac quantization for a quadratic constrained system.
We show that this solves the Jacobi identity violation problem occuring in the Dirac
quantization case and yields a well defined Fock space. 
By requiring the uniqueness of the ground state, we show that for non-constrained systems,
this approach gives the same results as Dirac quantization.
\end{abstract}
~~~After the formulation of quantum mechanics,
the passage from a classical dynamical system to a quantum one had to be properly
formulated. A solution  was to define the quantum mechanical
commutation relations in accordance with 
Poisson brackets. The commutator of operators ${\cal A}$ and ${\cal B}$
was set to be $i$ times the Poisson bracket of the classical quantities
$A$ and $B$. This works well for non-constrained systems. For
constrained systems, different approaches should be  adapted. Dirac's 
constraint theory \cite{dirac} deals with this problem using the Hamiltonian
approach. It first formulates new sets of brackets, namely Dirac brackets
(dirackets) using constraint equations and Poisson brackets. The passage to the
 quantum case is established via these dirackets. Recently, Jaroszkiewicz
\cite{george}
applied Dirac quantization to a constrained system described by a Lagrangian
in 5 dimensions
in order to obtain the operator algebra for quantized relativistic
space-time coordinates. \cite{snyder} He also showed that the operators
of this constrained system violate the  Jacobi identity. This result, referred as the 
Jacobi anomaly, was claimed by Dirac \cite{dirac} to be due to the fact 
that the process of taking
Poisson brackets does not commute with the process of applying constraints.\\

In this paper we will propose another way to pass from classical systems
to quantum systems, both for constrained and non-constrained systems. We will show
that, this approach is the same as canonical quantization for non-constrained
systems and it yields the Jacobi identity correctly for the exemplary case studied
in \cite{george}. The basic idea behind this approach is to relax the number of
commutation relations while passing from dirackets to quantum commutators. It will also
be shown that for this specific case, the quantum commutators
yield the Coon-Baker-Yu $q$ oscillator \cite{CBY} and a well defined positive definite
Fock space. \cite{arik_coon_lam} \\

The constrained system considered in \cite{george} has a  space-time  metric having the
signature $(+,-,-,-,-)$ or $(+,-,-,-,+)$ . We have found that, although we can solve the
Jacobi identity problem in both of these cases by using our basic idea, we could not solve
the unitarity problem due to the existence of a timelike coordinate after elimination of
the fifth coordinate using constraints.
Instead, by choosing the metric
as $(+,-,-,-,-)$ and eliminating the first coordinate using the same constraints we
have been able to construct, after quantization, a proper Hilbert space or in other words
a positive definite inner product.

The Lagrangian of this exemplary constrained system is given as:
\begin{equation}{\label{lagran}}
{\cal {L}} = {e \over 2}( {\dot Z}^2 + {\dot X}^{\mu}{\dot X}_{\mu} ) +
 {\alpha \over {2 e}} + \lambda ( Z^2 + X^{\mu}X_{\mu} - \beta)
\end{equation}
where $\mu = 1,2,3,4$, the 4-metric is $g_{\mu \nu}=-\delta_{\mu \nu}$. $\alpha$, $\beta$
are constants, and  $e \;$, $\lambda$ are parameters
which will be used to generate primary constraints. We will mainly be interested in the quantum
properties of this system. However after elimination of Lagrange parameters $e$ and $\lambda$, 
(\ref{lagran}) is clasically equivalent to the variational principle to obtain geodesics on the
pseudosphere $SO(4,1)/SO(3,1)$ for $\beta < 0$ and $SO(4,1)/SO(4)$ for $\beta >0$. The 
quantization scheme in not sensitive to the sign of $\beta$ and we prefer to work with the
first case since the tengent space in this case is the Minkowski space.
We also note that for $\lambda = 0$, (\ref{lagran})
is the parametrization invariant
Polyakov action \cite{poly} for a free particle in $4 + 1$ dimensions. 
 Here $Z$ refers to the zeroth dimension which is time-like. The
standard constraint analysis of Dirac yields:
\begin{eqnarray}{\label{pricon}}
\phi _1 &=& {P}^{\mu}{P}_{\mu} + {{\Pi ^2} 
     }-\alpha \;\approx \; 0 \; , \\
\phi _2 &=& X^{\mu}X_{\mu} + Z^2 - \beta \; \approx \; 0 \nonumber
\end{eqnarray}
as primary constraints, and
\begin{equation}{\label{seccon}}
\phi _3 = X^{\mu}P_{\mu} + Z \Pi \; \approx \; 0
\end{equation}
as the secondary constraint. Here $\Pi$ denotes the zeroth component of the momentum five-vector.
The Poisson brackets of these canonical quantities satisfy:
\begin{eqnarray}{\label{pb1}}
\{ P_{\mu}, X^{\nu} \} &=&  - \delta ^{\nu}_{\mu} \; , \\
\{ \Pi, Z \} &=& -1 \; . \nonumber 
\end{eqnarray}
For the classification of these constraints we evaluate their Poisson brackets and obtain:
\begin{eqnarray}{\label{pb2}}
\{ \phi _1, \phi _2 \} &=& 0 \; , \\
\{ \phi _1, \phi _3 \} &=& -2\alpha \; , \nonumber \\
\{ \phi _2, \phi _3 \} &=& 2 \beta \; . \nonumber 
\end{eqnarray}
The above equations show that we have one first class constraint and two second
class constaints. To ensure the compatibility with Jaroszkiewicz, we formulate
the second class constraints needed in the definition of dirackets as:
\begin{eqnarray}{\label{ksi}}
\chi _1 &=& f \phi _1 - g \phi _2 \; , \\
\chi _2 &=& \phi _3  \nonumber 
\end{eqnarray}
where $f$ and $g$ are constants satisfying,
 $h \equiv \alpha f + \beta g \; {\neq} \; 0 \;$.
Straightforward calculation of dirackets using $\chi _1$ and $\chi _2$ yields:
\begin{eqnarray}{\label{db1}}
\{ X^{\mu}, X^{\nu} \}_{DB} &=& - { f \over h} (X^\mu P^\nu - P^\mu X^\nu ) \; \\
\{ P^{\mu}, P^{\nu} \}_{DB} &=& - { g \over h} (X^\mu P^\nu - P^\mu X^\nu ) \; \nonumber \\
\{ P^{\mu}, X^{\nu} \}_{DB} &=& - \delta^{\mu \nu}+{ 1 \over h}(gX^\mu X^\nu +fP^\mu P^\nu )\nonumber
\end{eqnarray}
We note that the quantities on the right hand sides of the above equations are classical,
i.e. commuting ones. Thus they should be normalized in the conventional way while passing
to the quantum case. We also note that the left hand side of the first two equations are
antisymmetric and the right hand side of the third equation is symmetric under the 
interchange of $\mu$ and $\nu$.
The passage from dirackets to commutators is accomplished by replacing the
diracket of two momentum or two position variables by $-i$ times a commutator
whereas the diracket of a momentum and a position variable is replaced by
$i$ times a commutator, and all products on the right hand side are
symmetrized. After this procedure, the right hand side of the first
two equations
become antisymmetric whereas the right hand side of the third equation remains
symmetric under the interchange of $\mu$ and $\nu$. We write the resulting commutation equations
in terms of new constants
$A\;$, $B\;$, $C\;$ in order to allow rescaling of position and momentum coordinates:
\begin{eqnarray}{\label{prequ}}
\left[ X^\mu, X^\nu \right] &=& i {{-CB}\over {A}}( \left[ P^\mu, X^\nu \right]_+
-\left[ P^\nu, X^\mu \right]_+ )  \; \; , \\
\left[ P^\mu, P^\nu \right] &=& i {{-CA}\over {B}}( \left[ P^\mu, X^\nu \right]_+
-\left[ P^\nu, X^\mu \right]_+  )  \; \; ,  \nonumber \\
\left[ P^\mu, X^\nu \right] &=& -i \delta^{\mu \nu} + iC({A\over B} \left[ X^\mu,X^\nu \right]_+
+{B\over A} \left[ P^\mu,P^\nu \right]_+ )  \; \; , \nonumber 
\end{eqnarray}
where 
\[
{g\over {2h}} \equiv {{AC}\over B} \; \; \; ,  \; \; \; 
{f\over {2h}} \equiv {{BC}\over A} \; \; \; \hbox{and} \; \; \;
\left[ X,Y \right]_+ \equiv XY+YX \; \; .
\]
If we choose to represent our constrained quantum system with the above three equations, we
encounter the Jacobi anomaly problem. We believe that this failure is due to excess number
of commutation relations. Since we have investigated these three commutation equations
according to their behavior under the interchange of $\mu$ and $\nu$, we will try to group 
symmetric and antisymmetric equations separately. To accomplish this, firstly
we want to simplify our equations by eliminating some of
these constants and  we rescale $X$ and $P$ as follows:
\[
X \rightarrow {X \over A} \; \; \; \; \; \hbox{and} \; \; \; \; \; 
P \rightarrow {P \over B} \;  . \]
We can add, after rescaling, the first two equations of (\ref{prequ}) since they are both
antisymmetric. Meanwhile, the left hand side of the third equation has to be symmetrized
to keep the consistency with the right hand side. Thus we obtain two equations,
the first antisymmetric and the second symmetric:
\begin{eqnarray}{\label{quset}}
\left[X^\mu, X^\nu \right] +\left[P^\mu, P^\nu \right] &=& -i2C ( \left[ {P^\mu}, {X^\nu} \right]_+
-\left[ P^\nu, X^\mu \right]_+ )  \;  , \\
\left[P^\mu, X^\nu \right] +\left[P^\nu, X^\mu \right]  &=&
-i2AB \delta^{\mu \nu} + i2C(\left[X^\mu,X^\nu \right]_{+} +
 \left[ P^\mu,P^\nu \right]_{+} ) \; \; . \nonumber
\end{eqnarray}
These above equations constitute our basic system of quantization. Note that, we have reduced
the number of quantization equations from three to two.
In terms of components the first equation, antisymmetric under interchange of $\mu$ and $\nu$ contains $d(d-1)/2$ equations ($d = 4$) whereas the second, 
symmetric under interchange of $\mu$ and $\nu$ contains $d(d+1)/2$ equations
for hermitian operators. We call this, relaxing the number of commutations relations.
Moreover, if we define $C$ and $AB$ in terms of a new variable, $q$, as:
\begin{eqnarray}{\label{sltns}}
C &=& {{2(1-q)} \over {1+q}} \\
AB &=& {1 \over {1+q}} \nonumber
\end{eqnarray}
the above equations take the form:
\begin{eqnarray}{\label{qqset}}
\left[X^\mu, X^\nu \right] +\left[P^\mu, P^\nu \right] &=& -i \frac{1-q}{1+q}
\left( \left[ {P^\mu}, {X^\nu} \right]_+ -\left[ P^\nu, X^\mu \right]_+ \right) \; , \\
{ \left[P^\mu, X^\nu \right] +\left[P^\nu, X^\mu \right] } &=&
-i{2 \over {1+q}} \delta^{\mu \nu} 
+ i\frac{1-q}{1+q}(\left[X^\mu,X^\nu \right]_{+} + \left[ P^\mu,P^\nu \right]_{+} ) \nonumber
\end{eqnarray}
where $q \neq 1$. These two equations can be unified using creation and
annihilation operators defined through $a_\mu \equiv X_\mu+iP_\mu$. After
expressing (\ref{qqset}) in terms of $a$ and $a^\dagger$ we add symmetric and
antisymmetric parts to obtain a single compact equation:
\begin{equation}{\label{qosc}}
{{1-q} \over 2} \left[ a_\mu , {a^\dagger}_\nu \right]_{+} + {{1+q} \over 2} \left[ a_\mu , {a^\dagger}_\nu \right]
= \delta_{\mu \nu}
\end{equation}
which turns out to be same as the defining commutation relation of the
multidimensional Coon-Baker-Yu  $q$ oscillator algebra \cite{CBY}.
In fact, we define the
multidimensional CBY $q$ oscillator, in this metric, through the $q$ commutator
relation of $a$ and $a^\dagger$:
\begin{equation}{\label{qdef}}
\left[ a_\mu, a{^\dagger}{_\nu} \right]_q \equiv a_\mu a{^\dagger}_\nu -q
 a^{\dagger}_{\nu}a_{\mu}= \delta_{\mu \nu}
\end{equation}
which is exactly (\ref{qosc})  and which satisfies the Jacobi identity by definition. The
proof that such an algebra exists is the explicit construction of the Fock space associated
with this algebra  \cite{arik_coon_lam}. A positive definite scalar product requires
$ -1 < q < 1$.   

Since the examination of our basic idea showed us that for constrained systems,
quantization in Fock space is a generalization of the canonical (Dirac)
quantization, we now want to show that this can also be applied to non-constrained systems. Our starting point for quantization of non-constrained systems is
\begin{equation}{\label{osc}}
\left[ a_i, a^\dagger _j \right] = \delta_{ij}
\end{equation}
where the metric $\delta_{ij}$ is positive definite.
We want to stress that this defining  relation can be
decomposed in terms of  our symmetric and antisymmetric equations:
\begin{eqnarray}{\label{symasym}}
\left[ p_i ,q_j \right] + \left[ p_j ,q_i \right] &=& -i2\delta_{ij} \; \; ,\\
\left[ p_i ,p_j \right] + \left[ q_i ,q_j \right] &=& 0 \; \; . \nonumber
\end{eqnarray} 
Moreover in the case of a Hilbert space with a positive definite norm these
quantization rules give the same results as the canonical quantization case. This
can be proven by constructing the Fock space representation of oscillatory
states. We start by showing that the equations (\ref{symasym}) can be combined
to give (\ref{osc}). If annihilation and creation operators are defined as: 
\begin{equation}{\label{def}}
a_i \equiv {{q_i +i p_i} \over \sqrt{2}} \; \; \; , \; \; \;
a^\dagger_i \equiv {{q_i - i p_i} \over \sqrt{2}} \; \; , \nonumber
\end{equation}
we can calculate $\left[ a_i ,a^\dagger_j \right]$ directly to obtain
(\ref{osc}). With the assumption of the ground state, we can write
the action of any annihilation operator on this ground state to give zero,
i.e. $ a_i | \; \rangle \; = 0$ for any $i$. Then we will define oscillator
states as:
\begin{equation}{\label{state}}
| i_1 i_2 \cdots i_n \rangle \equiv a^\dagger_{i_1} a^\dagger_{i_2} \cdots
a^\dagger_{i_n} | \rangle \; \; \; .
\end{equation}
Using only (\ref{osc}), the matrix elements of $a^\dagger_i a^\dagger_j$ are given
by:
\begin{equation}{\label{inner}}
\langle k_1 k_2\cdots k_m | a^\dagger_i a^\dagger_j  | i_1 i_2\cdots i_n\rangle
= \delta _{m,n+2}N^{k_1 k_2 \cdots k_m}_{i j i_1 i_2 \cdots i_n} =
\sum^n_{l=1} \delta^{k_1}_{i_l} N^{k_2 \cdots \cdots \cdots k_m}_{i_1 \cdots i_{l-1} i_{l+1}
\cdots i_n}
\end{equation}
This expression will be completely symmetric under any permutation of the lower
or upper indices. This shows that
\[
\left[ a^{\dagger}_{j} , a^{\dagger}_{i} \right] = 0 \; \; \; \hbox {and thus} \; \; \;
 \left[ a_j , a_i \right] = 0 \; \; .
\]
Adding the last two equations we obtain $\left[p_i,p_j \right] = \left[q_i, q_j \right]$ . 
Combining these with (\ref{symasym}) we obtain the commutation relations in the
canonical quantization case as: \[ \left[p_i , p_j \right] = 0 \; \; , \; \; \left[ q_i , q_j \right] = 0 \; \;
\hbox{and} \; \; \left[ p_i, p_j \right] = -i \delta_{ij} \]
Note that, in the constrained case, the matrix element calculated above would
be the $q$ generalized version of the above $N$ tensor \cite{MMG}, which is not symmetric
under interchange of $i$ and $j$. Thus we cannot formulate any commutation
relations between $a_i$ and $a_j$; i.e.  canonical and non-canonical cases are
distinguished. Another point which also is worth noting is that, if one starts
with a d+1 dimensional model, a d dimensional oscillator is obtained
by a similar analysis. This model would yield a $q$ oscillator system
invariant under the classical group $U(d)$ which in the $q \rightarrow 1$ limit reduces
to the nonrelativistic $U(d)$ invariant quantum oscillator.

The underlying classical symmetry group was not modified during the
quantization scheme considered in this work. Instead, the number of commuation relations
were reduced
and the number of states were increased. However in the $q \rightarrow 1$ limit
($\hbar \rightarrow 0$ limit) the norms of these extra states became zero and
they are excluded from the Hilbert space \cite{MMG}. An alternative to this approach
is to change the symmetry
group itself and to require the invariance of commutation equations under
a quantum group. \cite{qgroup} The quantum group concept was first discovered through
the quantization of nonlinear, completely integrable models in two dimensions.
In the language of $q$ oscillators, a quantum group invariant
($U_q(n)$) system is given by Pusz-Woronowicz $q$ oscillators \cite{PW}.
Since a quantum group  is, in fact, mathematically not a group,
the concept of invariance under a classical group has to be generalized.
This depends on quantizing the Poisson brackets by using an exchange relation.
In this approach, the number of quantum commutators is equal to the number of
Poisson brackets leaving the number of states unchanged after the quantization.
Examples of this kind of quantization scheme were given in \cite{relpos}. If one
has to discover which quantization scheme nature has chosen,  a possibility
would be to set up an experiment to count the number of quantum states provided that the
measurements can be made to be sensitive to the states that arise due to the deviation
of the parameter $q$ from unity. The comparison of the number of states between the
non-canonical quantum system and its canonical
analogue would directly give us the preferred quantization scheme. 

We would like to thank George Jaroszkiewicz, Rufat M. Mir-Kasimov and
Vladimir I. Manko for discussions.

\end{document}